\ifdefined\manuscriptstyle\else\def\manuscriptstyle{manuscript}\fi
\ifdefined\manuscriptlayout\else\def\manuscriptlayout{traditional}\fi

\documentclass[journal=jcisd8,manuscript=article,layout=\manuscriptlayout]{achemso}
\mciteErrorOnUnknownfalse

\usepackage[version=3]{mhchem} % Formula subscripts using \ce{}
\usepackage[LGR,T1]{fontenc}   % Use modern font encodings
\usepackage{siunitx}
\usepackage{textcomp}
\usepackage{amsmath}
\usepackage{amssymb}
\usepackage{graphicx}
\usepackage{nicefrac}

\newcommand{\kbTe}{\ensuremath{k_\mathrm{B}T/|e|}}
\newcommand{\kbT}{\ensuremath{k_\mathrm{B}T}}
\DeclareSymbolFont{upgreek}{LGR}{cmr}{m}{n}
\DeclareMathSymbol{\epsilonup}{\mathord}{upgreek}{`e}

\newcommand\bioinfUDE{Bioinformatics and Computational Biophysics, Faculty of Biology, University of Duisburg-Essen, Universit{\"a}tstra{\ss}e 7, 45141 Essen, Germany}
\newcommand\orgUDE{Institute of Organic Chemistry, University of Duisburg-Essen, Universit{\"a}tstra{\ss}e 7, 45141 Essen, Germany}
\newcommand\crystUE{Laboratory of Chemical Biology, Department of Biomedical Engineering and Institute for Complex Molecular Systems, Eindhoven University of Technology, Den Dolech 2, 5612 AZ Eindhoven, The Netherlands}
\author{Jean-No\"{e}l Grad}
\affiliation[Bioinformatics and Computational Biophysics, Essen]{\bioinfUDE}
\author{Alba Gigante}
\affiliation[Institute of Organic Chemistry, Essen]{\orgUDE}
\author{Christoph Wilms}
\author{Jan Nikolaj Dybowski}
\author{Ludwig Ohl}
\affiliation[Bioinformatics and Computational Biophysics, Essen]{\bioinfUDE}
\author{Christian Ottmann}
\affiliation[Biomedical Engineering, Eindhoven]{\crystUE}
\author{Carsten Schmuck}
\affiliation[Institute of Organic Chemistry, Essen]{\orgUDE}
\author{Daniel Hoffmann}
\affiliation[Bioinformatics and Computational Biophysics, Essen]{\bioinfUDE}
\email{daniel.hoffmann@uni-due.de}
\phone{+49 (0)201 183 4391}
\fax{+49 (0)201 183 3437}

\title[Epitopsy]
  {Locating large flexible ligands on proteins%\footnote{A footnote for the title}
  }

\abbreviations{FFT}

\begin{document}

% \begin{tocentry}

% \includegraphics[height=1in]{figures/sonic/epi_occ_30p_xray.jpg}

% Computed region (green) of highest probability density of heparin di-saccharides around Sonic Hedgehog protein and crystallographic position of heparin tetra-saccharide (sticks).
% \end{tocentry}

\begin{abstract}
    \singlespace

Many biologically important ligands of proteins are large, flexible,
and often charged molecules that bind to extended regions on the protein
surface. It is infeasible or expensive to locate such ligands on proteins
with standard methods such as docking or molecular dynamics (MD) simulation.
The alternative approach proposed here is the scanning of a spatial
and angular grid around the protein with smaller fragments of the
large ligand. Energy values for complete grids can be computed efficiently
with a well-known Fast Fourier Transform accelerated algorithm and
a physically meaningful interaction model. We show that the approach
can readily incorporate flexibility of protein and ligand. The energy
grids (EGs) resulting from the ligand fragment scans can be transformed
into probability distributions, and then directly compared to probability
distributions estimated from MD simulations and experimental structural
data. We test the approach on a diverse set of complexes between proteins
and large, flexible ligands, including a complex of Sonic Hedgehog
protein and heparin, three heparin sulfate substrates or non-substrates
of an epimerase, a multi-branched supramolecular ligand that stabilizes
a protein-peptide complex, and a flexible zwitterionic ligand that
binds to a surface basin of a Kringle domain. In all cases the EG
approach gives results that are in good agreement with experimental
data or MD simulations.

\end{abstract}


    \singlespace

\section{Introduction}

The prediction of binding poses of small molecules with a mixture
of polar and hydrophobic groups that bind with high affinity in protein
pockets has been one of the dominating problems in biomolecular modeling,
and the successes in this endeavor had a major impact in the life
sciences and drug design. However, many biologically important interactions
are almost the exact opposite to this scenario: large, flexible ligands
bind to protein surfaces, their binding is often transient, and charge-charge
interactions are essential. Examples are interactions between secreted
proteins and the extracellular matrix of glycosaminoglycans\cite{Capila-2002,Coombe-2005},
interactions of virus proteins with host receptors in viral cell entry
\cite{Myszka2000}, or interactions of T-cell receptors with MHC
I-peptide complexes \cite{Willcox1999a}. Another interesting case
is the design of novel supramolecular ligands that bind protein surfaces
with many low affinity interactions but overall high avidity \cite{Gilles2017}.
How can we model and predict complexes of proteins with such large,
flexible ligands that are often charged or zwitterionic? Sometimes
it is possible to predict binding modes of large, flexible ligands
by docking suitable fragments with methods developed for small molecule
docking \cite{Jiang:2013}. This is less promising if binding occurs
not in typical small molecule binding pockets, but at the protein
surface, often involving charged residues with long, flexible side
chains, as for instance in the case of protein-glycosaminoglycan binding.
In these cases, interactions could be characterized by Molecular Dynamics
simulation (MD) or related sampling methods \cite{Yu2014}, though
the necessary computational effort can be excessive.

A promising alternative are approaches that evaluate energies for
ligand positions on a 3D-grid around the target protein. Although
they have mainly been used for docking \cite{Goodford1985}, i.e.~for
locating optimal ligand positions and poses, they allow in principle
for a characterization of the complete target protein surface and
environment with respect to ligand binding energetics. A great advantage
of grid-based approaches is that the protein-ligand interaction energies
on the grid can be evaluated efficiently by exploiting discrete Fast
Fourier Transforms (FFT) \cite{Katchalski-Katzir-1992,Gabb-1997,Kozakov-2006,Brenke-2009}.
Since we are mostly interested in interactions of proteins with charged
ligands, another candidate method for characterizing the interaction
energetics around the target protein is the solution of the Poisson-Boltzmann
equation, typically also with efficient grid-based methods \cite{Honig-1995,Baker-2001}.

In the work presented here we assess the suitability of fast grid-based
methods for predicting binding modes of large, flexible, and usually
charged ligands on protein surfaces. These ligands not only defy docking
methods, but they force us also to abandon the notion of the well-defined
binding pose, because their size and flexibility, and the fact that
they bind to extended protein surface regions will make binding more
fuzzy.

One way to account for this uncertainty while still retaining a quantitative
approach is to predict affinity distributions or probability densities
for the ligand, or at least for those functional groups that likely
mediate binding. The abovementioned grid-based methods \cite{Katchalski-Katzir-1992,Gabb-1997,Kozakov-2006,Brenke-2009}
are attractive because they could provide exactly this information
in an efficient way. Generally, the approach proposed here assumes
that we can infer the location of a large, flexible ligand from probability
distributions of characteristic fragments, and that these fragment
probability distributions can be computed efficiently and sufficiently
accurate by grid based, FFT accelerated scanning. We also demonstrate
that flexibility of the target protein and of ligand fragments can
be incorporated easily. 

To test the above assumptions we have applied the method to four different
test cases that cover several scenarios of practical interest: the
surface binding of heparin to Sonic Hedgehog protein for which we
compare several methods and experimental data, the specific interactions
of an epimerase with three different heparan sulfate substrates or
non-substrates as an example for specificity of interaction, the stabilization
of a protein-peptide complex by an artificial multi-branched supramolecular
ligand as example of a large non-polymeric ligand, and the binding
of a flexible zwitterionic ligand to a Kringle domain.

\section{Methods and molecules}

\subsection{Workflow}

\begin{figure}
\includegraphics[width=0.75\columnwidth]{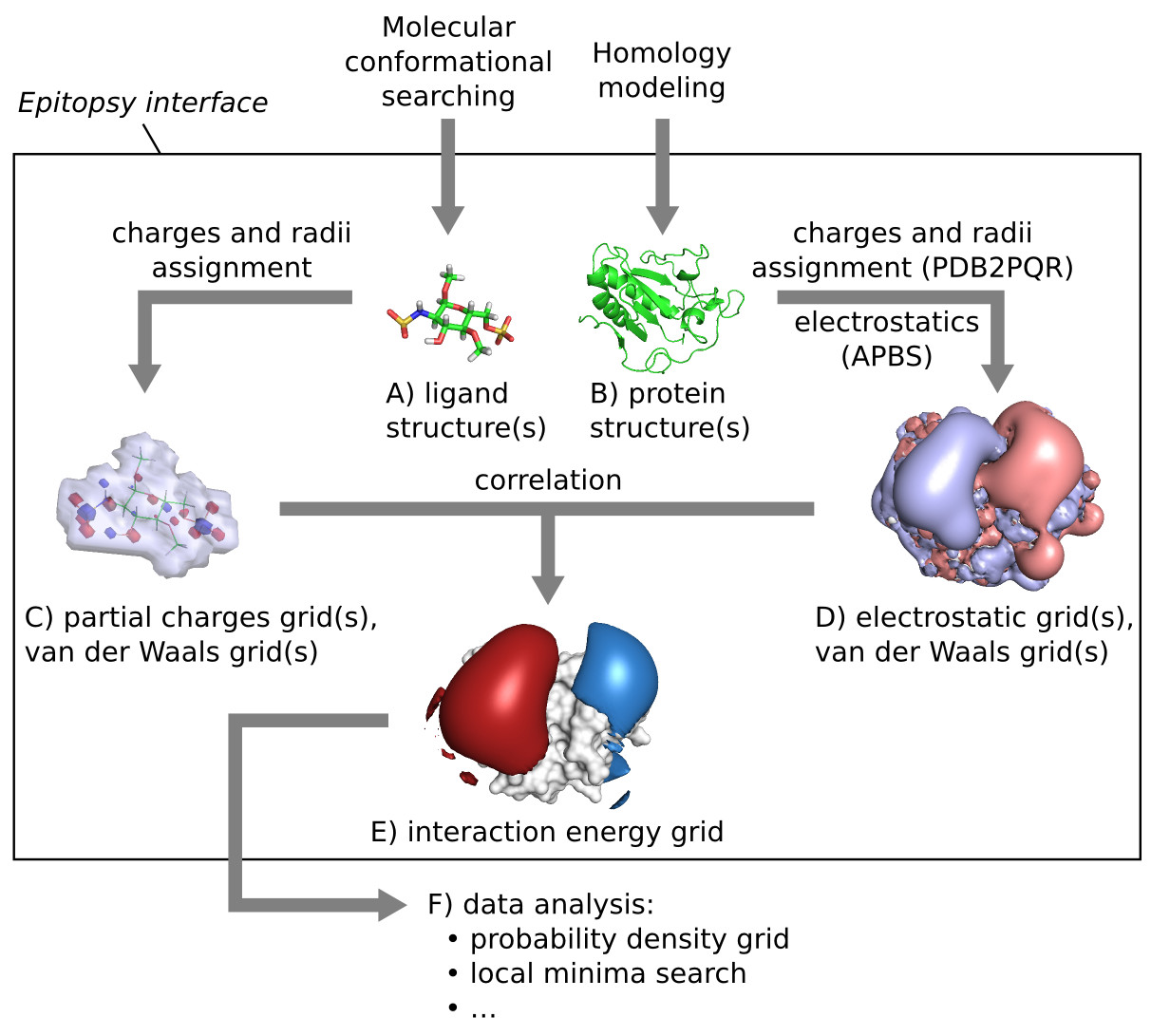}

\caption{Workflow for the computation of (interaction) energy grids (EGs) with
Epitopsy. Representative conformations of ligand fragments and target
protein are required as input (A, B). Charges and radii of both interaction
partners are assigned (C, D). The EG for a ligand-protein pair as
``correlation'' with FFTs as described in text (E). EGs can then
be analyzed in various ways outside Epitopsy (F). \label{fig:Workflow-for-the}}
\end{figure}

Figure~\ref{fig:Workflow-for-the} shows the overall workflow of
our grid-based analysis. Each of the steps will be described in the
following section. To make our results reproducible we provide our
experimental Epitopsy software as free open source code at\\ https://github.com/BioinformaticsBiophysicsUDE/Epitopsy.

\subsection{Molecular dynamics (MD) simulation}

MD simulations were used, first, to determine representative ligand
conformations as input of Epitopsy (Fig \ref{fig:Workflow-for-the}A),
and second, as a reference method to estimate probability densities
of ligand fragments around the target protein that can then be compared
to corresponding probabilities computed from (interaction) energy
grids (EGs) generated by Epitopsy.

Molecular Dynamics (MD) simulations were run with Gromacs 4.6.7\cite{Pronk-2013}
using the Amber ff99SB force field\cite{Hornak-2006} for proteins
and GLYCAM 06h-2\cite{Kirschner-2008} for saccharides. Phosphorylated
serine parameters were obtained from the literature\cite{Homeyer-2006}.
Non-standard amino acids in ligand QQJ-096 (succinic acid, phenyl
trihydrazine, N-acetyl-lysine and GCP) were parametrized for the ff99SB
force field according to the procedure described in the original ff94
article\cite{Cieplak-1995}. Atomic charges were derived from electrostatic
potential maps calculated at the HF/6-31{*}{*} level of theory in
Gaussian09 version A.02\cite{Frisch-2009a} and fitted to the residues
using the Restrained Electrostatic Potential method \cite{Bayly-1993,Cornell-1993}.
Force constant parameters were obtained by chemical analogy with readily
available parameters in ff94\cite{Cornell-1995}. Topology files
were created with the pdb2gmx module of Gromacs for the protein, and
with the TLEaP module of Amber v12.21\cite{Case-2012} with the AmberTools
suite v13.22 for the ligands. Amber topologies were converted to Gromacs
topologies by ACPype\cite{Silva-2012}.

Proteins and ligands were solvated in a dodecahedron box of SPC/E
water molecules\cite{Berendsen-1987} with a 10~{\AA} minimum separation
between the protein and the box boundaries. The system was neutralized
by addition of Na$^{+}$ and Cl$^{-}$ ions to a final ionic strength
of $0.15$~mol/l. The system was energy-minimized by steepest-descent
to a total force of $1800$~\si{kJ.mol^{-1}.nm^{-1}}, equilibrated
for 5~ns in the NVT ensemble with constrained heavy atoms, and for
5~ns in the NPT ensemble without constraints. Production simulation
was run in the NPT ensemble for 200~ns if not mentioned otherwise.
Temperature was stabilized at 300~K in the NVT and NPT ensembles
by the V-rescale thermostat\cite{Bussi-2007}, while the pressure
was stabilized at 1~atm in the NPT ensemble by the Berendsen barostat
(equilibration) or Parinello-Rahman barostat (data production)\cite{Parrinello-1981}.
Simulations were carried out on a GPU (GeForce 970 and GeForce 1070,
CUDA 8) using a time step of 2~fs, the Verlet scheme\cite{Pall-2013}
for neighbor search with a 10~{\AA} cutoff, the Particle Mesh Ewald
method\cite{Darden-1993} for electrostatic calculations, and the
LINCS algorithm\cite{Hess-1997} for bond restraints.

Representative structures were extracted from trajectories based on
mutual RMSDs, using the g\_rms tool in Gromacs to produce 2D RMSD
plots, the pam (partition around medoids) tool from R package \emph{cluster,}
version 2.0.6, in R v3.3.1\cite{RCoreTeam-2016} to find the clusters,
and the cluster.stats function of R package \emph{fpc,} version 2.1.10,
to validate the clustering based on silhouette coefficients. 

When high flexibility in the ligand prevented the extraction of representative
structures, the ligand trajectory was projected on a grid to produce
a probability distribution of the ligand around the protein. To this
end, the simulation box was discretized and we counted for each grid
point the number of MD frames where it was within the van der Waals
radius of a ligand atom. The resulting count was divided either by
the total number of frames in the trajectory to yield a grid point
sampling frequency, or by the sum of the grid point frequencies to
yield a (ligand) probability density. The latter was used to compute
cumulative density plots and to draw Highest Density Regions\cite{Hyndman-1996}
(HDR) in a molecular visualization software. When comparing electrostatic,
energy and probability density grids, all the compared grids were
laid out with the same resolution, dimensions and offset.

EGs, HDR and molecules were visualized with PyMOL v1.7.4.0\cite{PyMOL-2015}
compiled from sources.

\subsection{Protein structures}

Crystal structures were refined in Modeller 9.17\cite{Webb-2014,Eswar-2001}
to restore missing residues when necessary (Table~\ref{tab:protein-modeller}).
Candidate structures were required to minimize the DOPE and molpdf
score. In case of ties, the refined model with lowest RMSD to the
template was selected.

\subsection{Assignment of charges and radii}

For the computation of EGs, charges and radii have to be assigned
to the ligand (Fig \ref{fig:Workflow-for-the}C) and protein (Fig
\ref{fig:Workflow-for-the}D). Charges and atomic radii were added
on the proteins using PDB2PQR v2.0.0\cite{Dolinsky-2007,Dolinsky-2004}
at neutral pH and 298~K using the Amber force field option. Ligand
charges were determined with one of the following methods, as specified
in the text: with PDB2PQR (default), specialized MM forcefields, from
an electrostatic potential fit using the Merz-Singh-Kollman scheme\cite{Singh-1984,Besler-1990}
in Gaussian 2009 A02\cite{Frisch-2009a} at the HF/6-31G{*}{*} level
of theory, or using the Gasteiger-Marsili method\cite{Gasteiger-1978,Gasteiger-1980}
in OpenBabel v2.3.2\cite{OBoyle-2011}. Information on atomic radii
was added to the ligand atoms by OpenBabel.

\subsection{Electrostatics}

For the target protein the electrostatic field was computed by solving
the non-linear Poisson-Boltzmann equation with APBS version 1.4.1\cite{Baker-2001}
at 310K, with an ionic concentration of $0.15$~mol/l and relative
dielectric permittivities $\varepsilon_{r}^{\text{vacuum}}=2$ and
$\varepsilon_{r}^{\text{water}}=79$.

\subsection{Energy grid computation}

The central part of the workflow is the computation of the energy
grid (EG) for a ligand (or ligand fragment) and target protein (Figure~\ref{fig:Workflow-for-the}E).
As we are mainly interested in charged ligands, the energy model currently
only considers electrostatic interactions between ligand and protein
for non-overlapping relative positions and poses. EGs were calculated
using the EnergyGrid tool of Epitopsy 1.0\cite{Wilms-2013}. The
following subsections we describe how the energy is evaluated.

\subsubsection{Shape complementarity}

The atomic description of a protein \textendash{} obtained either
from experimentally solved structures or from homology modeling \textendash{}
is mapped to a grid of dimensions $(N_{1},N_{2},N_{3})$ with resolution
$(m_{1},m_{2},m_{3})$, usually in the range 0.5\textendash 1.0~{\AA}.
The default value in this work was 0.8~{\AA}. Discretization proceeds
by assigning a non-zero value to grid points within the van der Waals
radii defined by PDB2PQR for protein and ligand atoms. These discretized
geometries are labeled $\mathbf{f}_{\text{P}_{l,m,n}}^{\text{vdw}}$
for the protein and $\mathbf{f}_{\text{L}_{l,m,n}}^{\text{vdw}}$
for the ligand:

\begin{equation}
\begin{aligned}\mathbf{f}_{\text{P}_{l,m,n}}^{\text{vdw}} & =\begin{cases}
\delta & \text{protein}\\
+1 & \text{surface layer}\\
0 & \text{water}
\end{cases}\\
\mathbf{f}_{\text{L}_{l,m,n}}^{\text{vdw}} & =\begin{cases}
+1 & \text{ligand}\\
0 & \text{water}
\end{cases}
\end{aligned}
\label{eq:vdwbox}
\end{equation}

The surface layer is the ensemble of solvent grid points in direct
contact with the protein. The correlation is positive whenever the
ligand is in contact with the protein surface (i.e.~occupying the
surface layer), negative when the ligand overlaps the protein, and
zero otherwise. Ligand poses with negative shape correlation are discarded.
Flexibility is introduced by the use of coefficients with opposite
sign: an overlapping pose with $n$ overlapping grid points is rejected
unless a minimum of $|\delta\cdot n|$ grid points are in surface
contact. We used mainly $\delta=-15$ as given by \cite{Gabb-1997}
but point out in the discussion and Figure~\ref{fig:focusing} that
it can be useful to vary $\delta$.

The shape correlation $\mathbf{f}_{\text{C}_{\alpha,\beta,\gamma}}^{\text{vdw}}$
can be defined as the direct product of the two matrices $\mathbf{f}_{\text{P}_{l,m,n}}^{\text{vdw}}$
and $\mathbf{f}_{\text{L}_{l,m,n}}^{\text{vdw}}$ for any shift vector
$(\alpha,\beta,\gamma)$: 

\begin{equation}
\mathbf{f}_{\text{C}_{\alpha,\beta,\gamma}}^{\text{vdw}}=\sum\limits _{l}^{N_{1}}\sum\limits _{m}^{N_{2}}\sum\limits _{n}^{N_{3}}\mathbf{f}_{\text{P}_{l,n,m}}^{\text{vdw}}\mathbf{f}_{\text{L}_{l+\alpha,n+\beta,m+\gamma}}^{\text{vdw}}\label{eq:vdw-corr-direct}
\end{equation}

This calculation has an asymptotic time complexity of $O\left(n^{6}\right)$,
making it impractical for solving numerically large systems. The Fast
Fourier Transform (FFT) $\mathcal{F}$ (or $\mathcal{F}^{-1}$ for
the reverse operation) was successfully introduced by Gabb \emph{et
al.} \cite{Gabb-1997} in this context, resulting in a time complexity
of $O\left(n^{3}\ln\left(n^{3}\right)\right)$:

\begin{equation}
\begin{aligned}\mathbf{F}_{\text{P}}^{\text{vdw}} & =\mathcal{F}\left\{ \mathbf{f}_{\text{P}_{l,m,n}}^{\text{vdw}}\right\} \\
\mathbf{F}_{\text{L}}^{\text{vdw}} & =\mathcal{F}\left\{ \mathbf{f}_{\text{L}_{l,m,n}}^{\text{vdw}}\right\} \\
\mathbf{F}_{\text{C}}^{\text{vdw}} & =\overline{\mathbf{F}_{\text{P}}^{\text{vdw}}}\mathbf{F}_{\text{L}}^{\text{vdw}}\\
\mathbf{f}_{\text{C}_{\alpha,\beta,\gamma}}^{\text{vdw}} & =\mathcal{F}^{-1}\left\{ \mathbf{F}_{\text{C}}^{\text{vdw}}\right\} 
\end{aligned}
\label{eq:vdw-corr-FFT}
\end{equation}

where the uppercase letter $\mathbf{F}$ represents the decomposed
signal $\mathbf{f}$ , and $\overline{\mathbf{F}}$ is the complex
conjugate of $\mathbf{F}$ .

\subsubsection{Electrostatic energy}

The electrostatic potential (ESP) $\mathbf{\Phi}_{l,m,n}$ of the
protein in ionic aqueous solution obtained by solving the non-linear
Poisson-Boltzmann equation (see above) is stored in a matrix $\mathbf{\mathbf{\Phi}}_{\text{P}_{l,m,n}}$,
with the protein interior and surface set to a potential of zero.
The matrix $\mathbf{q}_{\text{L}_{l,m,n}}$ contains the ligand partial
charges: 
\begin{equation}
\begin{aligned}\mathbf{\mathbf{\Phi}}_{\text{P}_{l,m,n}} & =\begin{cases}
0 & \text{protein}\\
\mathbf{\Phi}_{l,m,n} & \text{water}
\end{cases}\\
\mathbf{q}_{\text{L}_{l,m,n}} & =\begin{cases}
\mathbf{q}_{l,m,n} & \text{ligand}\\
0 & \text{water}
\end{cases}
\end{aligned}
\label{eq:espbox}
\end{equation}

The electrostatic interaction between the ligand partial charges and
the protein electrostatic potential is used to compute the energy
correlation matrix $\mathbf{f}_{\text{C}_{\alpha,\beta,\gamma}}^{\text{elec}}$:

\begin{equation}
\mathbf{f}_{\text{C}_{\alpha,\beta,\gamma}}^{\text{elec}}=\sum\limits _{l}^{N_{1}}\sum\limits _{m}^{N_{2}}\sum\limits _{n}^{N_{3}}\mathbf{\Phi}_{\text{P}_{l,n,m}}\mathbf{q}_{\text{L}_{l+\alpha,n+\beta,m+\gamma}}\label{eq:esp-corr-direct}
\end{equation}

The same FFT optimization described in Equation~\ref{eq:vdw-corr-FFT}
is used to speed up the correlation here:

\begin{equation}
\begin{aligned}\mathbf{F}_{\text{P}}^{\text{elec}} & =\mathcal{F}\left\{ \mathbf{\Phi}_{\text{P}_{l,m,n}}\right\} \\
\mathbf{F}_{\text{L}}^{\text{elec}} & =\mathcal{F}\left\{ \mathbf{q}_{\text{L}_{l,m,n}}\right\} \\
\mathbf{F}_{\text{C}}^{\text{elec}} & =\overline{\mathbf{F}_{\text{P}}^{\text{elec}}}\mathbf{F}_{\text{L}}^{\text{elec}}\\
\mathbf{f}_{\text{C}}^{\text{elec}} & =\mathcal{F}^{-1}\left\{ \mathbf{F}_{\text{C}}^{\text{elec}}\right\} 
\end{aligned}
\label{eq:esp-corr-FFT}
\end{equation}

\subsubsection{Correlation matrix}

The energy correlation matrix $\mathbf{f}_{\text{C}_{\alpha,\beta,\gamma}}^{\text{elec}}$
(Equation~\ref{eq:vdw-corr-direct}) is the electrostatic contribution
$\mathbf{\Delta E}_{\text{bind}}^{\text{elec}}(\alpha,\beta,\gamma)$
to the binding affinity for any shift vector $(\alpha,\beta,\gamma)$
where the molecular probe does not overlap with the protein, i.e.~for
$\mathbf{f}_{\text{C}_{\alpha,\beta,\gamma}}^{\text{vdw}}\geq0$:

\begin{equation}
\mathbf{\Delta E}_{\text{bind}}^{\text{elec}}(\alpha,\beta,\gamma)=\mathbf{f}_{\text{C}_{\alpha,\beta,\gamma}}^{\text{elec}}H\left[\mathbf{f}_{\text{C}_{\alpha,\beta,\gamma}}^{\text{vdw}}\right]\label{eq:G-discrete}
\end{equation}

with Heaviside step operator 

\begin{equation}
\begin{aligned}H[x]=\begin{cases}
1, & x\geq0,\\
0, & x<0
\end{cases}\end{aligned}
\label{eq:Heaviside}
\end{equation}

\subsubsection{Angular sampling and energy grid values}

\label{sub:Rotations}

The correlation matrices are evaluated for many orientations $\mathbf{\omega}\in\mathbf{\Omega}$
of the ligand, where $\mathbf{\Omega}$ is a set of $(\phi,\theta)$
tuples uniformly distributed on a sphere using a Fibonacci generative
spiral \cite{Swinbank-2006,Gonzalez-2009}. The binding free energy
$\mathbf{\Delta G}_{\text{bind}}^{\text{elec}}(\alpha,\beta,\gamma)$
is computed from $|\mathbf{\Omega}|$ correlations:

\begin{equation}
\mathbf{\Delta G}_{\text{bind}}^{\text{elec}}(\alpha,\beta,\gamma,\mathbf{\Omega})=-\kbT\ln\left(\dfrac{\sum\limits _{\mathbf{\omega}\in\mathbf{\Omega}}\exp\left(-\mathbf{f}_{\text{C}_{\alpha,\beta,\gamma}^{\mathbf{\omega}}}^{\text{elec}}H\left[\mathbf{f}_{\text{C}_{\alpha,\beta,\gamma}^{\mathbf{\omega}}}^{\text{vdw}}\right]\right)}{|\mathbf{\Omega}|}\right)\label{eq:G-discrete-rotations}
\end{equation}

The division by $|\mathbf{\Omega}|$ accounts for the purely entropical
free energy of the reference state, i.e.~the ligand immersed in pure
solvent where it takes $|\mathbf{\Omega}|$ orientations of zero enthalpy.
We used mainly $|\mathbf{\Omega}|=150$ as it provides a reasonable
trade-off between accuracy and calculation time; we show in Figure~\ref{fig:angular-sampling}
the effect of increasing $|\mathbf{\Omega}|$.

The number of available ligand rotations at every grid point is

\begin{equation}
\mathbf{\Omega}^{\text{available}}(\alpha,\beta,\gamma,\mathbf{\Omega})=\sum\limits _{\mathbf{\omega}\in\mathbf{\Omega}}H\left[\mathbf{f}_{\text{C}_{\alpha,\beta,\gamma}^{\mathbf{\omega}}}^{\text{vdw}}\right]\label{eq:G-discrete-rotations-microstates}
\end{equation}

The ligand excluded volume (LEV) corresponds to the set of grid points
$\alpha,\beta,\gamma$ where no rotational state with finite energy
is available to the ligand ($\mathbf{\Omega}^{\text{available}}=0$),
or the set of all grid points where function $\mathbf{LEV}$ is 1:

\[
\mathbf{LEV}(\alpha,\beta,\gamma,\mathbf{\Omega})=1-H\left[\mathbf{\Omega}^{\text{available}}-1\right]
\]

\subsubsection{Energy grids for multiple conformers}

\label{sub:Conformations}

When several conformers $\mathbf{P}$ of the protein and $\mathbf{L}$
of the ligand are provided, with respective internal energies $U_{i}$
for $\mathbf{P}$ and $U_{j}$ for $\mathbf{L}$, the binding free
energy is:

\begin{equation}
\mathbf{\Delta G}_{\text{bind}}^{\text{elec}}(\alpha,\beta,\gamma,\mathbf{P},\mathbf{L},\mathbf{\Omega})=-\kbT\ln\left(\dfrac{\sum\limits _{i\in\mathbf{P}}e^{-U_{i}}\sum\limits _{j\in\mathbf{L}}e^{-U_{j}}\sum\limits _{\mathbf{\omega}\in\mathbf{\Omega}}\exp\left(-\mathbf{f}_{\text{C}_{\alpha,\beta,\gamma}^{ij\mathbf{\omega}}}^{\text{elec}}H\left[\mathbf{f}_{\text{C}_{\alpha,\beta,\gamma}^{ij\mathbf{\omega}}}^{\text{vdw}}\right]\right)}{|\mathbf{\Omega}|\cdot\sum\limits _{i\in\mathbf{P}}e^{-U_{i}}\sum\limits _{j\in\mathbf{L}}e^{-U_{j}}}\right)\label{eq:G-discrete-conformers}
\end{equation}

The number of available orientations is:

\begin{equation}
\mathbf{\Omega}^{\text{available}}(\alpha,\beta,\gamma,\mathbf{P},\mathbf{L},\mathbf{\Omega})=\sum\limits _{i\in\mathbf{P}}\sum\limits _{j\in\mathbf{L}}\sum\limits _{\mathbf{\omega}\in\mathbf{\Omega}}H\left[\mathbf{f}_{\text{C}_{\alpha,\beta,\gamma}^{ij\mathbf{\omega}}}^{\text{vdw}}\right]\label{eq:G-discrete-conformers-microstates}
\end{equation}

\subsubsection{Conversion to probability densities}

EGs may be transformed into probability density functions (PDFs) with
Boltzmann factors $\mathbf{K}$ for positions outside the LEV:

\[
\mathbf{K}(\alpha,\beta,\gamma)=\left(1-\mathbf{LEV}(\alpha,\beta,\gamma,\mathbf{\Omega})\right)\exp\left(\dfrac{-\mathbf{\Delta G}_{\text{bind}}^{\text{elec}}(\alpha,\beta,\gamma,\mathbf{\Omega})}{\kbT}\right)
\]

\begin{equation}
\mathbf{PDF}(\alpha,\beta,\gamma)=\dfrac{\mathbf{K}(\alpha,\beta,\gamma)}{\text{\ensuremath{\sum\mathbf{K}}}(\alpha,\beta,\gamma)}\label{eq:PDF}
\end{equation}

\section{Results}

\subsection{Sonic Hedgehog and heparin}

The complex of Sonic Hedgehog protein (Shh) and a heparin ligand is
prototypical for our systems of interest: a large, flexible, and highly
charged ligand binds to the surface of a protein. The general assumption
underlying our computational assessment of the heparin location is
that we can infer the location of large, flexible ligands from the
probability densities of characteristic fragments. Of course this
assumption has to be tested, and it will break down under certain
conditions as we outline in the discussion. To test the approach we
have therefore compiled for this system a comprehensive data set,
consisting of the electrostatic potential (ESP) of Shh, energy grid
(EG) for a di-saccharide heparin fragment scan of Shh, seven 500~ns
MD simulations of these di-saccharide fragments with Shh (sufficiently
long to observe ligand binding and unbinding events, Figure~\ref{fig:shh-heparin-unbinding}),
and the crystal structure of the Shh-heparin tetra-saccharide complex
from \cite{Whalen-2013}. 

\begin{figure*}[!hb]
\includegraphics[width=1\linewidth]{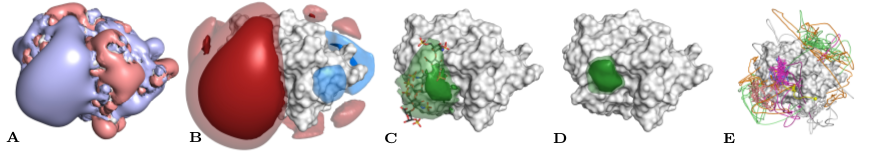} \caption{Sonic Hedgehog protein (Shh) with heparin ligand. (A) ESP isosurfaces
of Shh at $+1$~$\kbTe$ (blue) and $-1$~$\kbTe$ (red). (B) EG
isosurfaces at $\pm1$~$\kbT$ (translucent blue/red) and $\pm2$~$\kbT$
(solid blue/red), merging across two population-weighted conformations
of a heparin di-saccharide (clustering details and glycosidic angles
are provided in Table~\ref{tab:gags-topologies}) according to Equation~\ref{eq:G-discrete-conformers}.
(C) EG based probability density of heparin di-saccharide drawn around
20\% HDR (solid green) and 30\% HDR (translucent green). The 20\%
HDR forms a hull around the crystallographic position of the heparin
tetra-saccharide from \cite{Whalen-2013} (PDB entry 4c4n).(D) MD
based probability density of heparin di-saccharide drawn at 20\% and
30\% HDR (solid and translucent green) from a 3.5~\textmu s multi-trajectory
MD simulation. (E) MD traces of the seven 500~ns simulations.}

\label{fig:shh-heparin-esp-epi-occ} 
\end{figure*}

Overall, the four sources of data give a consistent picture (Figure~\ref{fig:shh-heparin-esp-epi-occ}):
the ESP has its largest high-potential region around Arg156, and this
is where the EG has its largest low energy blob, and where the heparin
tetra-saccharide is located in the crystal structure, and this is
also the area of the highest heparin di-saccharide probability density,
as estimated from MD trajectories. 

\begin{figure}[!ht]
\centering \includegraphics[width=0.75\linewidth]{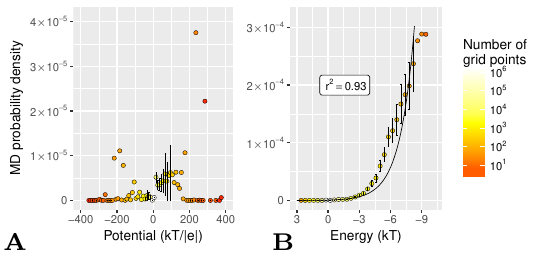}
\caption{Probability density of heparin di-saccharide occupancy computed from
MD simulations vs.~electrostatic potential (A) and energy grid (B)
at the same grid positions. Each point in the plot stands for all
grid points with a certain value of potential (A) or energy (B), as
given by its position along the horizontal axes. The number of grid
points with the respective ESP or EG values are shown as colors. Both
horizontal axes go from repulsive to attractive, and in both panels
the vertical axes give the probability estimated by MD sampling, averaged
over the grid points with a given ESP or EG value. The error bars
mark the 99\% confidence interval, assuming normally distributed probabilities.
In panel A, thirteen outliers in the ESP grid with energies ranging
from 400 to 800~$\kbTe$ lie outside the plotting range. Figure~\ref{fig:shh-heparin-occ-esp-epi-2d-hist}
shows the distribution in a 2D histogram.}

\label{fig:shh-heparin-occ-esp-epi-exp} 
\end{figure}

Assuming a Boltzmann distribution, the EG and ESP values can be transformed
into probability densities for ligand occupancy (Equation~\ref{eq:PDF}).
This probability density can then be compared directly with the probability
density estimated by MD sampling, either visually, e.g.~with 3D-isosurfaces
(Figure~\ref{fig:shh-heparin-esp-epi-occ}), or quantitatively (Figure~\ref{fig:shh-heparin-occ-esp-epi-exp}).
For the latter we evaluated the frequencies with which the heparin
di-saccharide visited each EG/ESP grid cube in the concatenated MD
trajectories as described in Methods. The highest probability density
from EG and MD is located in the same area around Arg156 (Figure~\ref{fig:shh-heparin-esp-epi-occ}C,D)
where the EG shows its by far largest low energy blob (Figure~\ref{fig:shh-heparin-esp-epi-occ}B).
However, the EG probability density maximum there has a much larger
spatial spread than the MD probability density. Interestingly, the
region of 20\% highest probability density as computed from heparin
di-saccharide EGs forms an envelope around the crystal position of
the heparin tetra-saccharide, following the crystal ligand in shape
and size (Figure~\ref{fig:shh-heparin-esp-epi-occ}C). This supports
our initial hypothesis that we can locate larger ligands from probability
distributions of fragments.

In a more quantitative comparison between EG and MD probability densities
(Figure~\ref{fig:shh-heparin-occ-esp-epi-exp}B) we see that the
MD probability density roughly follows an exponential of the EG values,
as expected for a Boltzmann distribution (coefficient of determination
$r^{2}=0.93$). The deviation between the actual distribution and
an exponential could be a result of unequilibrated MD sampling or
of EG model deficiencies.

The obvious similarity of ESP and EG (Figure~\ref{fig:shh-heparin-esp-epi-occ}A,B)
suggests that ESP should have a similarly good association with MD.
However, this is not the case (Figure~\ref{fig:shh-heparin-occ-esp-epi-exp}A).
If we transform ESPs into probability densities for a charged ligand,
the probability density is almost completely concentrated at a single
grid point close to the two-calcium center of Shh, 2~nm away from
Arg156. While this point is certainly very attractive for the heparin
di-saccharide if we only consider Coulomb interactions, it is sterically
not accessible and therefore neither visible in the EG nor sampled
by MD. Figure~\ref{fig:shh-heparin-occ-esp-epi-exp}A suggests that
the same is true for many points of high ESP that are barely explored
in MD simulations or evaluated in the EG.

\subsection{C$_{5}$-epimerase and poly-anionic heparan sulfate substrates and
non-substrates}

{\scriptsize D}-glucuronyl C$_{5}$-epimerase modifies heparan sulfate
(HS), i.e.~long, negatively charged, and highly flexible carbohydrate
chains. The epimerase has a varied surface topography with deep clefts.
The HS chains have to be threaded through a narrow, partially buried
active site, which makes the epimerase-HS complex a harder test case
than the Shh-heparin complex of the previous section, where heparin
bound preferentially to a well-accessible surface patch on Shh. The
more specific, conformation dependent chemical function of the epimerase
suggests a more accurate positioning of the HS substrate chains on
epimerase than the superficial attachment of heparin to Shh. The hypothesis
of a more accurate positioning is consistent with the observed substrate
length dependency of the reaction: enzymatic activity decreased by
90\% on a digested heparan sulfate fraction containing octasaccharides
and smaller oligosaccharides \cite{Jacobsson-1984}. Our question
was therefore whether we would be able to trace an extended binding
site in EGs that could accommodate such longer oligosaccharides. For
validation we compared the predicted binding sites with crystallographically
determined binding sites with a heparin inhibitor (PDB entry 4pxq
\cite{Qin-2015}). 

\begin{figure*}[!ht]
\includegraphics[width=1\linewidth]{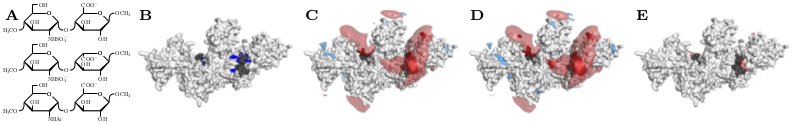} \caption{{\scriptsize D}-glucuronyl C$_{5}$-epimerase interaction with heparan
sulfate. (A) di-saccharides used to compute EGs around epimerase.
Top: substrate CH$_{3}$O-GlcNS-GlcA-OCH$_{3}$; middle: product and
\emph{in vitro }substrate CH$_{3}$O-GlcNS-IdoA-OCH$_{3}$; bottom:
non-substrate CH$_{3}$O-GlcNAc-GlcA-OCH$_{3}$. (B) Crystal structure
of epimerase in complex with heparin hexamer (PDB entry 4pxq \cite{Qin-2015}).
The two heparin fragments (black) bind at the two active sites of
the C$_{2}$ symmetric enzyme dimer. Amino acids critical for reaction
(Ala-mutations lead to enzyme activity loss of >60\% compared to wild-type
\cite{Qin-2015}) are marked in blue. C--E: EGs of substrate (C),
product (D), and non-substrate (E) scanned using the apo protein (PDB
entry 4pw2 \cite{Qin-2015}) with isosurfaces drawn at $-1$~$\kbT$
(translucent red) and $-2$~$\kbT$ (solid red) and crystallographic
heparin (black space filling). Isosurfaces in C--E were robust against
changes of dihedral angles of the HS dimer used for scanning. Glycosidic
angles are provided in Table~\ref{tab:gags-topologies}.}

\label{fig:glucuronyl-C5-epimerase-epi}
\end{figure*}

We used three different HS dimer fragments (Figure~\ref{fig:glucuronyl-C5-epimerase-epi}A)
to compute the EGs: CH$_{3}$O-\-GlcNS-\-GlcA-OCH$_{3}$ as model of the
substrate, CH$_{3}$O-GlcNS-\-IdoA-OCH$_{3}$ as model of the product,
and CH$_{3}$O-GlcNAc-\-GlcA-OCH$_{3}$ as a non-substrate \cite{Jacobsson-1984,Lindahl-1989}.
Note that \emph{in vitro} the enzyme works both ways, i.e.~the product
is a substrate for the reversed reaction \cite{Lindahl-1989}. A
parsimonious, natural explanation of this finding is that substrate
and product use the same molecular binding site.

In fact, in the EGs with epimerase substrate and product we detected
the same low energy channel, centered around the active sites (Figure~\ref{fig:glucuronyl-C5-epimerase-epi}B--D).
The region that binds most strongly in the EG matches the crystallographic
positions of the heparin hexasaccharide, and covers the amino acid
residues most important for enzymatic activity \cite{Qin-2015}.
However, the low energy region extends noticeably beyond the crystallographic
location of the heparin hexasaccharide and could easily accommodate
HS oligomers longer than octasaccharides (translucent red in Figure~\ref{fig:glucuronyl-C5-epimerase-epi}C,D).
The shape of this low energy region suggests a core binding site for
HS chains reaching from the right flank of the narrow cleft with the
active center down the crystallographic heparin binding site. 

While the substrate and product are both doubly negatively charged,
the non-substrate molecule (bottom of Figure~\ref{fig:glucuronyl-C5-epimerase-epi}A)
carries only one negative charge. In the corresponding EG the $-1$~$\kbT$
region has shrunk drastically and now only covers the location of
the crystallographic heparin hexasaccharide. Thus, although the non-substrate
could be chemically epimerized in principle -- it has the same GlcA
amenable to epimerization -- this particular epimerase enzyme offers
no suitable binding site for a longer chain of this non-substrate
type.

\subsection{14-3-3 protein and poly-cationic supramolecular ligand}

\begin{figure*}[!ht]
\includegraphics[width=1\textwidth]{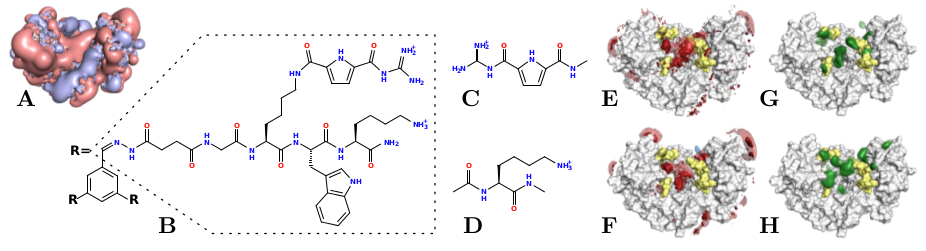}\caption{Interaction of 14-3-3/c-Raf complex with supramolecular ligand QQJ-096.
(A) isosurfaces of electrostatic potential at $+1$~$\kbTe$ (blue)
and $-1$~$\kbTe$ (red) of 14-3-3/c-Raf (PDB entry 4ihl \cite{Molzan-2013}).
(B) QQJ-096 ligand with only one of three arms (``R'') shown. (C)
GCP with capped ends. (D) Lys with capped ends. (E and F) EG computed
with Epitopsy for GCP and Lys, respectively, around 14-3-3/c-Raf.
(G and H) 20\% HDR (solid green) and 30\% HDR (translucent green)
for GCP and Lys, respectively, from a 1.5~\textmu s multi-trajectory
MD simulation of the 14-3-3/c-Raf with QQJ-096 in aqueous solution.}

\label{fig:14-3-3-esp-epi-occ}
\end{figure*}

Recently we could demonstrate experimentally (Gigante \emph{et al}.,
unpublished) that the binding of a supra{\-}molecular ligand, QQJ-096\cite{Jiang-2015}
(Figure~\ref{fig:14-3-3-esp-epi-occ}B), stabilizes the interaction
between the 14-3-3 protein and peptide fragments of c-Raf protein
(we call this complex 14-3-3/c-Raf). The large QQJ-096 ligand has
three flexible arms (``R'' in Figure~\ref{fig:14-3-3-esp-epi-occ}B),
each of them ending in two positively charged groups, a Lysine and
Guanidinocarbonylpyrrole (GCP). The size and flexibility of the ligand
makes it unsuitable for small molecule docking, and it is unlikely
that this ligand takes a single, well-defined binding pose. Nevertheless
its effect on the interaction of c-Raf and 14-3-3 could be explained
most easily by a specific binding of QQJ-096 to 14-3-3/c-Raf. 

The electrostatics of 14-3-3/c-Raf shows many of regions of low electrostatic
potential (red in Figure~\ref{fig:14-3-3-esp-epi-occ}A) that could
interact with the positive end groups of QQJ-096, Lys and GCP. For
a more ligand-specific assessment of binding, we computed two EGs
with molecules corresponding to the end groups of QQJ-096, GCP (Figure~\ref{fig:14-3-3-esp-epi-occ}C)
and Lys (Figure~\ref{fig:14-3-3-esp-epi-occ}D). The EGs show roughly
the same features for GCP (Figure~\ref{fig:14-3-3-esp-epi-occ}E)
and Lys (Figure~\ref{fig:14-3-3-esp-epi-occ}F), with particularly
high affinities in the center of the 14-3-3 cleft between the c-Raf
peptides. For Lys there are additional high affinity patches so that
the c-Raf fragments are sandwiched between regions of high affinity
of Lys. Both end groups have a few small high affinity islands outside
the central cleft of 14-3-3/c-Raf, with Lys having more of those islands
than GCP. Overall, this result suggests that QQJ-096 could seal off
the 14-3-3/c-Raf cleft and in this way inhibit dissociation of c-Raf
fragments, in agreement with experimental results (Gigante \emph{et
al.,} unpublished).

For comparison we simulated the 14-3-3/c-Raf/QQJ-096 system with MD.
Considering the size of the molecular system and the low charge density
on the ligand, a computational experiment analogous to the Shh-heparin
experiment above seemed to be unfeasible, i.e.~we do not expect to
reach the MD steady state in the microsecond time scale with a ligand
initially positioned at random in the solvent box. Based on the experimental
evidence for a QQJ-096-mediated stabilization of the 14-3-3/c-Raf
complex, and assuming a direct mode of interaction, the MD starting
conditions can be narrowed down to the 14-3-3 cleft (Figure~\ref{fig:input-QQJ-096}).
Based on this reasoning, we carried out a pilot set of six 50~ns
MD simulations of the 14-3-3/c-Raf dimer in aqueous solution with
QQJ-096 initially positioned 10~{\AA} above the c-Raf peptides.
In three simulations the ligand failed to interact with the protein.
We then ran six 250~ns simulations with the ligand initially positioned
4--6~{\AA} above the c-Raf peptides and observed a quick convergence
to binding sites of QQJ-096 end groups in the 14-3-3 cleft matching
those predicted by EGs computed with the end groups (Figure~\ref{fig:14-3-3-esp-epi-occ}G,H).
Regions outside the cleft were barely explored. Thus, MD simulations
and EGs both support the same mechanism for the experimentally observed
stabilization of the 14-3-3/c-Raf binding by QQJ-096, namely that
the supramolecular ligand QQJ-096 blocks the 14-3-3 cleft and in this
way impedes escape of c-Raf.

\subsection{Kringle domain and flexible zwitterionic ligand}

\begin{figure*}[!hb]
\includegraphics[width=1\linewidth]{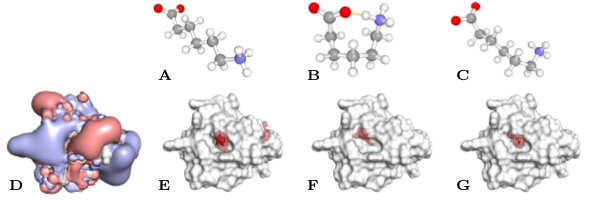} \caption{$\epsilonup$-aminocaproic acid (EACA) and its complex with plasminogen
Kringle domain 4 (KR4). (A) Experimental structure (``stretched'')
of EACA in the solid phase (CCDC entry 1102509\cite{Bodor-1967}).
B: Energy-minimized structure of EACA (``turn'', MMFF94 forcefield,
steepest descent in vacuum). (C) Structure of EACA in complex with
KR4 (PDB entry 2pk4). (D) Electrostatic potential isosurfaces around
apo KR4 (PDB entry 2pk4) drawn at $+1$~$\kbTe$ (blue) and $-1$~$\kbTe$
(red). E-G: EG isosurfaces a drawn at $\pm1$~$\kbT$ (translucent
blue/red) and $\pm2$~$\kbT$ (solid blue/red) of apo KR4 with EACA
stretched (E), as turn (F), and merging across multiple conformations
of the ligand and protein according to Equation~\ref{eq:G-discrete-conformers}
(G). }

\label{fig:plasmin-esp-epi2} 
\end{figure*}

The Kringle domains of plasminogen attach to Lys residues on fibrin,
a precondition for the decomposition of fibrin by plasminogen. A known
alternative ligand of the Kringle domains is $\epsilonup$-aminocaproic
acid (EACA), and a crystal structure of its complex with plasminogen
Kringle domain 4 (KR4) has been determined (PDB entry 2pk4 \cite{Wu1991}).
EACA is a highly flexible, zwitterionic molecule (Figure~\ref{fig:plasmin-esp-epi2}A--C)
that binds to a shallow basin in the KR4 surface. We have used the
EACA-KR4 complex as a test case for the application of EGs based on
multiple ligand conformers for the identification of binding sites
(Equation~\ref{eq:G-discrete-conformers}). In an application scenario
we would probably not know the actual conformer but rely on plausible
ligand conformers obtained from other experiments or modeling. Accordingly,
our EACA input conformers were the stretched conformer(Figure~\ref{fig:plasmin-esp-epi2}A)
observed in the solid phase of pure EACA \cite{Bodor-1967}, and
a MM energy minimized turn geometry (Figure~\ref{fig:plasmin-esp-epi2}B)
that is entropically and enthalpically more favorable for a free ligand.
The conformer actually observed in the crystal complex (Figure~\ref{fig:plasmin-esp-epi2}C)
is closer to the stretched geometry in the solid phase (Figure~\ref{fig:plasmin-esp-epi2}A),
though with a bent amino-end.

The zwitterionic nature of EACA suggests a binding site that bridges
two regions of opposite electrostatic potential. However, this pattern
is too unspecific since there are many regions that fall into this
category (Figure~\ref{fig:plasmin-esp-epi2}D). The full correlation
with shape and electrostatics information leads to the identification
of the correct binding basin, that, in fact, bridges regions of opposite
electrostatic potential (Figure~\ref{fig:plasmin-esp-epi2}E--G).
For the stretched EACA conformer there are two binding sites, the
one in the crystal structure, and an alternative binding site with
slightly lower affinity between Asp381 and Lys433 (Figure~\ref{fig:plasmin-esp-epi2}E).
For the EACA turn conformer the correct basin is clearly the region
with highest affinity (Figure~\ref{fig:plasmin-esp-epi2}F). The
EG averaged over both ligand conformers (Equation~\ref{eq:G-discrete-conformers},
both ligand conformations weighted equally) also has the basin of
the crystal structure as clearly dominating binding site (Figure~\ref{fig:plasmin-esp-epi2}G).

A question related to the multiconformer ligand treatment is the multiconformer
receptor treatment, and Equation~\ref{eq:G-discrete-conformers}
treats ligand and receptor symmetrical in this respect. In fact, Figure~\ref{fig:plasmin-esp-epi2}G
is based not only on two ligand conformations but also on three, equally
weighted KR4 receptor conformations, including the EACA-KR4 complex
structure 2pk4, a KR4 complex with sulfate (PDB entry 1krn), and a
KR4 complex with arginine (PDB entry 4duu). However, since the differences
between receptor structures are small (RMSDs to 2pk4: $0.29$~{\AA}
for 1krn and $0.69$~{\AA} for 4duu on $\text{C}_{\alpha}$ atoms)
the results are very similar for the combined EG and for the three
EGs based on single receptor structures: For all three apo KR4 structures
the respective EG identified the same correct EACA binding basin as
best binding site.

\section{Discussion}

The very nature of large, flexible ligands, such as glycosaminoglycan
chains, large receptor loops, or novel supramolecular binders, makes
it challenging to model their interactions with proteins. First, these
ligands are too large and flexible for small-molecule docking. Even
the notion of a well-defined binding pose, commonly used in small-molecule
docking, probably has to be abandoned. Instead, we should restate
the aim from finding the binding pose to computing a probability density
for the ligand around the protein. Second, the vastness of their conformational
space makes large, flexible ligands also difficult objects for the
standard method MD simulation, as it will be prone to severe undersampling
\cite{Neale2014,Nemec2017}. A third established candidate method
is continuum electrostatics. It takes advantage of the charged nature
of the ligands, but the electrostatic potential alone is difficult
to interpret. The fact that electrostatics alone is not sufficient
to locate large charged ligands like DNA or RNA has been noted before
\cite{Jones-2003,Chen-2008}. Our results corroborate this because
we found only a weak correlation between the electrostatic potential,
the probability density obtained from extensive MD simulations of
the Shh-heparin system, and corresponding experimental data. In this
case we achieved good consistency with MD simulation and experimental
structures with EGs that complement electrostatics with information
about shape and volume of the ligand or ligand fragments.

The good correlation of the EG approach with probability densities
from extensive MD simulations at a small fraction of the computational
cost (typically CPU minutes vs.~weeks) makes EGs an interesting way
to approximate such densities. However, there are also limitations
that should be considered. There are two basic categories of deficiencies:
first, those due to features of the real system that are missing in
the model, and second, those due to inadequate configuration of the
model. 

Intra-ligand interactions fall into the first category of deficiencies.
In the fragment-based screening used for the EGs, we neglect interactions
between fragments. This can be problematic since we are looking at
large, flexible ligands that carry charges, and that can have plenty
of opportunities for such interactions, e.g.~repulsion between same
sign charges, salt-bridges, or $\pi$-cation interactions. It is clear
that such interaction exist, and that they can have an impact on the
ligand structure\cite{Lavery-2014}. A factor that could limit the
severity of the effect of intra-molecular interactions is that they
have to compete with ligand-protein and ligand-water interactions,
and with the conformational entropy of the ligand. 

Another example of a principal deficiency of the underlying model
is the use of continuum electrostatics. This leads e.g.~to the neglect
of structural water molecules, although such waters can be crucial
for specific interactions at the surface of charged proteins \cite{Materese-2009,Davey-2002}.
A similar argument can be made for ions. Inclusion of fixed water
molecules or ions in the protein structure is technically feasible
in EG computations. For a first tests we used the Trp repressor interaction
with a nucleotide ligand where structural water molecules are known
to mediate interactions \cite{Otwinowski1988}. In this case the
presence of the known structural water molecules had a negligible
effect on the ligand distribution (Figure~\ref{fig:explicit-waters}).

Yet another deficiency is the treatment of molecular flexibility.
In the case of the Kringle domain we have demonstrated that flexibility
can be included both on the side of the protein and the side of the
ligand fragments. However, the relevant conformations have to be known
beforehand. If the protein-ligand interaction induces a new set of
conformations, the distributions obtained from the EGs may be misleading.
While this has not been an issue in the cases discussed above, it
could be relevant for highly flexible or disordered proteins.

Finally, the FFT based correlation computation used in Epitopsy treats
protein and ligand inconsistently: while the protein is modeled as
uniform medium with a low dielectric constant, the ligand is treated
as point charges in continuum water. For flexible charged ligands
this approximation can be acceptable because they will be well-solvated
and polarizable, but it could become inaccurate for ligands with a
more rigid structures or non-polar regions.

The second category of deficiencies can be controlled by proper configuration
of the model. For instance, if the protein or the ligand fragment
used for the computation of the EG can assume several drastically
different conformations and the user chooses only one of those conformations,
this will in general lead to systematic deviations between the computed
and true densities. This problem can be solved by inclusion of several
conformations (Equation~\ref{eq:G-discrete-conformers}).

Another source of errors that can be controlled is the grid configuration.
If spatial or angular grids are too coarse, regions around the protein
that contribute to the EG will be missed. In the present study we
have used fragments with the maximum size of a di-saccharide and a
grid spacing of 0.8~{\AA}, and the comparison with MD and experimental
data showed that the results are reasonable for the given ligands
and proteins. However, we expect that problems will arise with increasing
ligand fragment size and ruggedness of protein topography (Figure~\ref{fig:ligand-size}).
For instance, the larger the ligand fragment and the deeper protein
pockets, the more difficult it will be to map the protein-ligand interaction
on the angular and spatial grid, because many fragment poses will
lead to collisions and therefore be discarded. Another useful parameter
in this context is the clash penalty $\delta$ (Equation~\ref{eq:vdwbox}).
A weaker penalty will increase noise but has also the potential of
making visible finer structures in EG or probability density (Figure~\ref{fig:focusing}).

The correct charge of the ligand fragment is crucial, as shown in
the epimerase example. However, the approach is robust against small
variations in the charge distribution (see e.g.~Figure~\ref{fig:ligand-charge-distribution}),
so that resource-intensive QM-based methods for charge assignment
may be substituted with MM forcefield charges.

An important point that has not been addressed in this work is heterogeneous
ligand composition. In the presented examples we could infer the location
of larger ligands from fragment probability densities because the
large ligand had a rather homogeneous composition, e.g.~it was a
heparin poly-saccharide with negative charges on all di-saccharides,
or a multi-branched ligand with positively charged GCP and Lys groups.
However, such large ligands may comprise subunits of different physico-chemical
characters. In this case information of EGs for different ligand fragments
have to be combined to infer likely locations of complete ligands.
We are currently developing methods to post-process sets of EGs for
heterogeneous ligands in this sense.

We have argued in the beginning that current computational methods
are not suitable for the treatment of large, flexible ligands or that
their application is very expensive. Unfortunately, the same applies
also to high-resolution experimental characterization by X-ray crystallography
or NMR. This makes it all the more important to develop reliable and
efficient computational methods that can e.g. be used to predict protein
residues that are crucial for the interaction with the ligand. Such
predictions can then be tested e.g.~by measuring affinity changes
after site-directed mutagenesis.

\section{Author contribution}

Conceived and designed the experiments: JNG DH. Performed the calculations:
JNG. Analyzed the data: JNG LO DH. Contributed computational tools:
JNG CW JND LO. Contributed experimental data: AG CS CO. Wrote the
paper: JNG DH.

\section{Competing interests}

The authors of this manuscript have read the journal policy and declare
that they have no competing interests.

\begin{acknowledgement}
This work was supported by Deutsche Forschungsgemeinschaft through
grant CRC 1093 to CS and AG (subproject A1), JNG, LO, DH (subproject
A7), and CO (subproject B4). 
\end{acknowledgement}

\section*{Supporting information}
%\begin{suppinfo}
\setcounter{figure}{0}
\renewcommand{\thefigure}{S\arabic{figure}}
\setcounter{table}{0}
\renewcommand{\thetable}{S\arabic{table}}
\setcounter{equation}{0}
\renewcommand{\theequation}{S\arabic{equation}}
\refstepcounter{table}Table~\thetable\label{tab:gags-topologies}: Geometries of the glycosaminoglycans used as input for EGs. \refstepcounter{figure}Figure~\thefigure\label{fig:input-QQJ-096}: {Initial placement of QQJ-096 in the MD simulations with 14-3-3/c-Raf}. \refstepcounter{table}Table~\thetable\label{tab:protein-modeller}: Refined PDB structures. \refstepcounter{figure}Figure~\thefigure\label{fig:angular-sampling}: {Effect of the number of rotations $\left|\mathbf{\Omega}\right|$}. \refstepcounter{figure}Figure~\thefigure\label{fig:focusing}: {Effect of the grid resolution and penalty $\delta$}. \refstepcounter{figure}Figure~\thefigure\label{fig:explicit-waters}: {Effect of explicit water molecules}. \refstepcounter{figure}Figure~\thefigure\label{fig:ligand-size}: {Effect of the ligand size}. \refstepcounter{figure}Figure~\thefigure\label{fig:ligand-charge-distribution}: {Effect of the ligand charge distribution}. \refstepcounter{figure}Figure~\thefigure\label{fig:shh-heparin-unbinding}: {Binding/unbinding events in the heparin/Sonic Hedgehog MD
simulations}. \refstepcounter{figure}Figure~\thefigure\label{fig:shh-heparin-occ-esp-epi-2d-hist}: {2D histograms of the ESP and EG vs. MD probability density}. 
%\end{suppinfo}

\providecommand{\latin}[1]{#1}
\providecommand*\mcitethebibliography{\thebibliography}
\csname @ifundefined\endcsname{endmcitethebibliography}
  {\let\endmcitethebibliography\endthebibliography}{}

\end{document}